\documentclass[useAMS,usenatbib]{mn2e}

\usepackage{graphicx}

\title[Observing CMB polarisation through ice]{Observing CMB polarisation through ice}

\author[Luca Pietranera et al.]
        {Luca Pietranera$^{1,3}$\thanks{E-mail: luca.pietranera@astro.cf.ac.uk},  Stefan A. Buehler$^{2}$, Paolo G. Calisse$^{3}$, Claudia
        Emde$^{4}$, \and Darren Hayton$^{3}$, Viju Oommen John$^{5}$, Bruno Maffei$^{1}$, Lucio
        Piccirillo$^{1}$, \and Giampaolo Pisano$^{1}$, Giorgio Savini$^{3}$, T. R. Sreerekha$^{6}$\\
$^{1}$School of Physics and Astronomy, University of Manchester, Manchester, M13 9PL UK\\
$^{2}$Institut für Umweltphysik, Universit$\ddot{a}$t Bremen, Bremen, 28359, Deutschland\\
$^{3}$School of Physics and Astronomy, Cardiff University Cardiff, CF24 3AA, UK\\
$^{4}$Deutsches Zentrum fuer Luft- und Raumfahrt (DLR) Institut fuer
Physik der Atmosphaere, Oberpfaffenhofen, D-82234 Germany\\
$^{5}$Rosenstiel School of Marine and Atmospheric Sciences, University of Miami,USA\\
$^{6}$Met Office, Exeter, EX1 3PB, UK}

\begin{document}

\date{Accepted YYYY Mmmmmm dd. Received YYYY Mmmmmm dd; in original form YYYY Mmmmmm dd}

\pagerange{\pageref{firstpage}--\pageref{lastpage}} \pubyear{2002}

\maketitle

\label{firstpage}

\begin{abstract}
Ice crystal clouds in the upper troposphere can generate
polarisation signals at the $\umu$K level. This signal can seriously
affect very sensitive ground based searches for E- and B-mode of
Cosmic Microwave Background polarisation. In this paper we estimate
this effect within the C$_{\ell}$OVER experiment observing bands
(97, 150 and 220 GHz) for the selected observing site (Llano de
Chajnantor, Atacama desert, Chile). The results show that the
polarisation signal from the clouds can be
 of the order of or even bigger than the CMB expected polarisation.
 Climatological data suggest that this signal is fairly constant over the whole year in Antarctica.
 On the other hand the stronger seasonal variability in Atacama allows for a
 50\% of clean observations during the dry season.

\end{abstract}


\section{Introduction}

C$_{\ell}$OVER (C$_{\ell}$ObserVER) is a collaboration between the
Cardiff Astronomy Instrumentation Group, Oxford Astrophysics,
Manchester Astrophysics and the Cavendish Astrophysics Group in
Cambridge, on an experiment to measure the Cosmic Microwave
Background (CMB) polarisation \citep{b22}.

Polarisation of the CMB is caused by Thomson scattering of CMB
photons at the last scattering surface \citep{b19}. The signal can
be decomposed into a curl and a curl-free component, known as B- and
E-mode. The B-mode signal, which is at best one order of magnitude
weaker than the E-mode, is generated by primordial tensor
perturbations and therefore its detection would provide valuable
information about the history of the early universe.

The main ambitious objective of C$_{\ell}$OVER is the measurement of
the B-mode; in order to achieve this result the experiment will
deploy large format imaging arrays, operating at 97, 150 and 220 GHz
with 30\% bandwidth and a beamwidth of approximately 8 arcmin; the
instrument  is designed with an unprecedented level of systematic
control and will be deployed in the Atacama desert (Chile) at an
altitude of 5080 m. Alternative sites are the Antarctic stations of
Dome C and South Pole.

In spite of a site choice with favorable atmospheric conditions we
expect the signal of the atmospheric fluctuations to be well above
the intrinsic instrumental noise.

Since the C$_{\ell}$OVER receiver modulates signal polarisation, the
main concern about atmospheric effects is about a potentially
polarised signal from the atmosphere. Water vapor is the major
absorbing component at mm wavelengths and its spatial distribution
is highly variable with time.

These variations could also introduce some polarisation noise; in
situ measurements of the turbulence suggest that this polarised
contribution to system noise is expected to be gaussian and
negligible during most of the observing time, even in Atacama which
should be the worst of the three sites, both because of the stronger
day-night thermal cycle and the height of the mean boundary layer
that can be from 200 to 2000 meters \citep{b15}, with respect to 230
m in South Pole and 30 m in Dome C \citep{b1} during winter.

In addition to the variable contribution by water vapor, the strong
oxygen features at 120 GHz and around 60 GHz dominate the brightness
temperature of the atmosphere in C$_{\ell}$OVER's spectral region.
The presence of the Earth's magnetic field causes a Zeeman splitting
of the energy levels of the oxygen, thus resulting in a polarised
emission depending on the relative alignment between the line of
sight and the magnetic field. This effect is well known for
atmospheric measurements (see \citet{b10}, and references cited
therein). \citet{b21} discussed its impact on CMB measurements.
\citet{b17} estimated that the circularly-polarised component is not
negligible if the intrinsic leakage between linear and circular
polarisation in the instrument will be of the order of one per cent.

However, the polarised intensity due to oxygen is not expected to
vary with time but is be fixed for a particular azimuth and
elevation direction. Hence any scanning strategy will modulate any
residual atmospheric signal in a very predictable way. Also, the
estimates are for the DC level of the signal. Oxygen is well mixed
in the atmosphere at altitudes up to approximately 80 km, hence
fluctuations in the oxygen signal on the angular scales to which
C$_{\ell}$OVER is sensitive will be very small. It should therefore
be possible to separate this signal from the CMB polarisation well
down below the sensitivity required.

Upper tropospheric ice clouds (like cirrus clouds) represent another
source of polarised radiation. These clouds are at high altitudes
and contribute to the energy budget of the atmosphere (greenhouse
effect) since they absorb thermal IR radiation from the ground and,
as they are cold, emit little infrared radiation. This warms up the
Earth-atmosphere system. On the other hand, ice clouds reflect
incoming solar short wave radiation and hence cool the
Earth-atmosphere system.

At mm and sub-mm wavelengths the interaction between ice clouds and
radiation is mainly due to scattering. Absorption is negligible, and
so is the thermal emission. The scattering by ice clouds will
introduce a polarisation signal. \citet{b30} have shown that this
polarisation signal arises even assuming spherical ice particles,
due to the asymmetry of the radiation field in the atmosphere.
However, real cloud ice particles are not spherical \citep{b33}, and
this increases the polarisation signal. Moreover, there is a growing
evidence of horizontal alignment of cloud ice particles due to a
combination of aerodynamic and gravitational forces \citep{b27},
which further increases the polarisation signal. The actual
magnitude of the cloud polarisation signal will depend strongly on
the particle size and shape, and on the line of sight direction.

Experimentally, ice crystal depolarisation is a well-known problem
for high frequencies satellite telecommunications (20$\div$50 GHz)
based on signal polarisation diversity encoding \citep{b23}.
Measurements carried out with experimental telecommunication
payloads (ITALSAT and OLYMPUS) showed that even at relatively low
frequencies (with respect to C$_{\ell}$OVER bands) the depolarising
effect of ice crystals is not negligible \citep{b31}.

For CMB measurements, the impact of the cloud scattering is twofold.
Firstly, the CMB signal is depolarised, similarly to a
telecommunication signal.  Secondly, the cloud also scatters back
upwelling thermal radiation from the earth surface into the line of
sight of the instrument.  For telecommunication links this second
effect is negligible, due to the large intensity of the
telecommunication signal.  But for CMB measurements the radiation
scattered back by the atmosphere will often be more intense than the
CMB signal.  The backscattered signal will be partially polarised,
and its polarisation characteristics will depend on many factors, as
will be explained in the following sections.

\begin{table}
 \centering
\caption{Candidate C$_{\ell}$OVER observation sites.}
  \begin{tabular}{@{}cccc@{}}
  \hline
     Site name  &  Latitude ($^\circ$) & Longitude ($^\circ$) & Height (m) \\

 \hline
 Atacama & 23 S & 67 W & 5080 \\
 Dome C & 75 S & 123 E & 3280 \\
 South Pole & 90 S & - & 2900 \\
 \hline
\end{tabular}
\end{table}

\begin{figure*}
  % \vspace*{280pt}
\includegraphics [width=18 cm, clip=true]{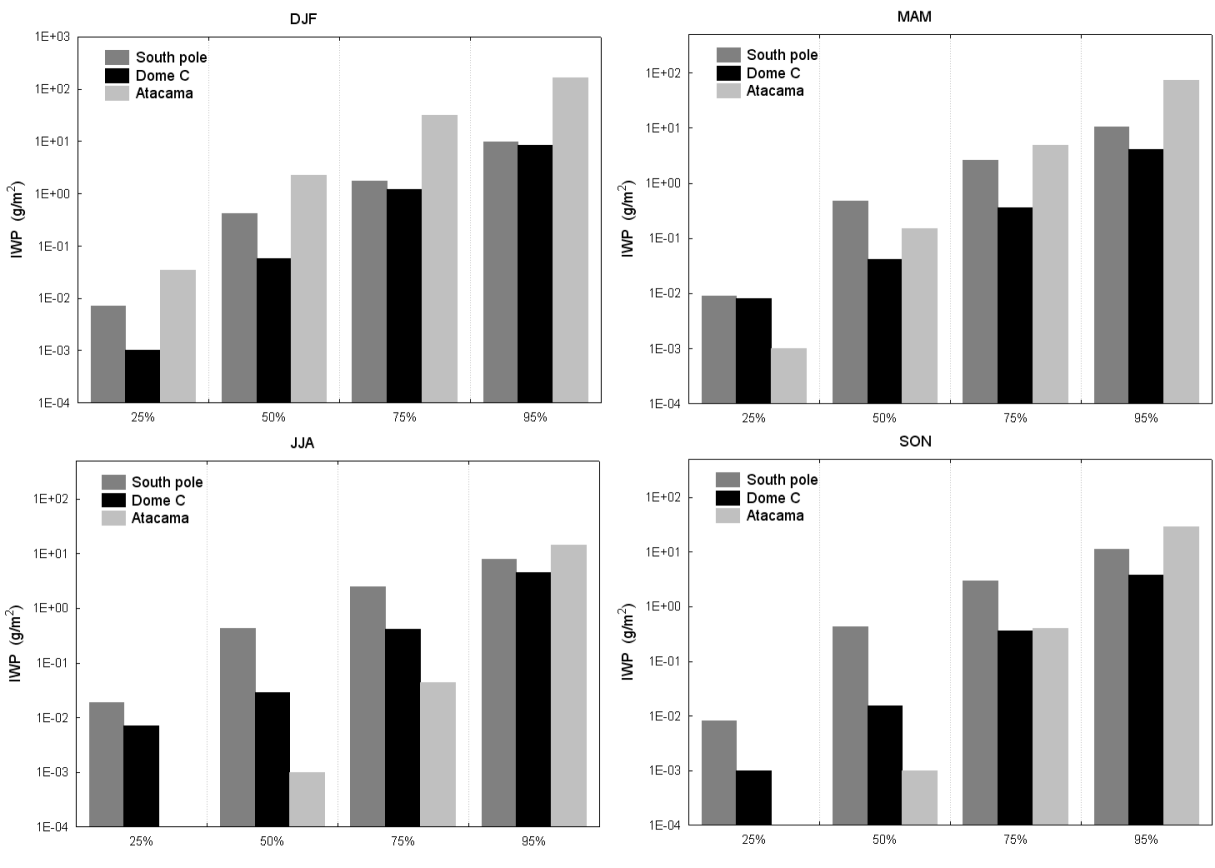}\\
\caption{IWP percentiles (values express column density, units are
g/m$^2$); each graph represents a three month period. IWP values for
the 25\% percentile in Atacama, during JJA and SON, are $\leq$
0.0001 g/m$^2$.}
\end{figure*}

\section[]{Ice in the upper troposphere  \\* for the three sites}

Although ice clouds play an important role in the atmosphere energy
budget, up to now they are poorly measured and modeled.

Satellites provide global measurements of integrated ice mass (Ice
Water Path - IWP) with frequent revisit time on a long term basis.
Sensors detect both reflected sunlight (in the UV and visible,
\citet{b5}) and thermal emission. While the last method is limited
to semitransparent clouds the first one only works if the earth
surface albedo is not too high (which, unfortunately, is exactly the
case for Atacama desert and Antarctica).

The MODIS instrument carried by the Terra and Aqua EOS-NASA
satellites is equipped with a cirrus clouds detection band at 1.38
$\umu$m; the method, first suggested by \citet{b14} for airborne
measurements, suffers the same drawback since it assumes that
upwelling radiation reflected by the earth surface is strongly
absorbed by water vapor in the lower troposphere and therefore the
method is not effective when water vapor column density is very low
(i.e. $<$ 4 kg/m$^2$ corresponding to 4 mm precipitable water vapor,
PWV).

The next generation satellite or air-borne instrument will
characterize ice clouds by measuring from above the radiation
brightness temperature depression with imaging radiometers in the mm
and sub-mm range, as proposed by \citet{b3} and \citet{b13} and will
include also polarisation measurements \citep{b16}. In situ and
aircraft-borne experiments provide the most accurate information on
ice clouds. Dual polarisation radars (30-90 GHz), polarisation
diversity lidars and airplanes equipped with cameras are used to
characterize ice density, crystal shapes, orientation and size
distribution.

As mentioned above, there is evidence that ice needles and plates
(especially those with large size and aspect ratio) have a preferred
orientation. As crystals drift downwards, they become oriented in a
maximum drag condition: aerodynamic forces tend to cause their long
axis (or axes) to fall horizontally (i.e. the shortest axis is
perpendicular to the ground; see \citet{b12} and references therein
for both models and some experimental results). Such a good
orientation is testified also by the relatively frequent presence of
optical effects, such as sun haloes, which happens only if the
crystal are aligned within few degrees. More recently, \citet{b26}
derived this result from polarised lidar backscatter measurements.
\citet{b25} analysed polarised visible light measurements from the
POLDER satellite, and found that 50\% of high clouds show a glint
signature implying at least a fraction of the ice platelets to be
horizontally aligned (within a very narrow angle). \citet{b6} showed
that cloud ice crystals generate a polarisation signal in the limb
measurements at 122 GHz (carried out by the Microwave Limb Sounder
(MLS) on the Aura satellite). They also concluded that the effective
particle shape can be approximated by a horizontally oriented oblate
spheroid with an aspect ratio of 1.3. This rather moderate value of
the effective asphericity is due to the fact that, while there are
individual particles with large aspect ratios, there is also an
averaging effect over the different sizes, shapes and orientations
of the individual ice particles.

The purpose of this paper is to estimate the influence of cloud ice
particles on the C$_{\ell}$OVER measurements at the selected and at
the two backup sites (Table 1). For this purpose assumptions are
made on the range of cloud ice amount to be expected for the
different sites, as well as on the ice particles size, shape, and
orientation.

We used the general circulation model of the European Centre for
Medium Range Weather Forecasting (ECMWF, see \citet{b32}) to
estimate the statistics of cloud ice mass. Figure 1 shows the
statistics in the form of percentiles. It shows for example that
Atacama, though generally a very dry place, can have ice clouds
exceeding an Ice Water Path (i.e. ice column density, IWP hereafter)
of 100 g/m$^2$ for about 10\% of the time from December to May.
These statistics were derived from a three years long (2000 to 2002)
data set obtained with a gridded version of the ECMWF model data
with a grid resolution of $1.5^{\circ}$ by $1.5^{\circ}$. A longer
data set is still under analysis, however we believe that the period
considered is quite significant since it was not affected by
climatic extremes (such as El Niño events in South America).

\begin{figure}
  \vspace*{0pt}

\includegraphics [width=8 cm]{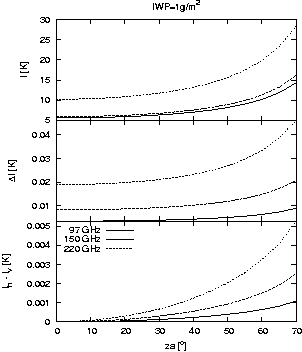}\\
\caption{Simulated cloudy radiances, radiance difference and
polarisation difference between a cloudy and a clear sky as a
function of instrument zenith angle.
   Different curves are for different observation frequencies.}
\end{figure}

As expected, Atacama shows a much stronger seasonal variability than
Antarctica, in the driest season IWP is often (25\% of time) $\leq$
0.0001 g/m$^2$. Since this is the selected site for C$_{\ell}$OVER
experiment, we evaluated the polarised signal from ice crystals
using a standard atmospheric profile for these latitudes, re-scaled
in accordance with locally measured water vapor values (1 mm PWV,
i.e. 1 kg/m$^2$).

The impact on CMB polarisation measurements will be discussed for a
wide range of IWP values from 0.0001 g/m$^2$ to 100 g/m$^2$.

\begin{figure}
 \vspace*{0pt}

  \includegraphics [width=8 cm]{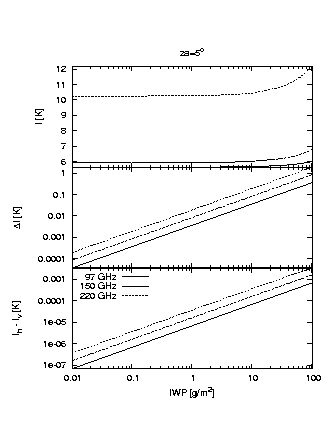}\\
    \caption{Simulated radiance, radiance difference and polarisation
  difference between a cloudy and a clear sky at a zenith angle of  5$^\circ$
  as a function of IWP.}
  \end{figure}

\section{Scattering model}

There are no direct data available for particle size, shape and
orientation. A realistic size distribution based on literature was
adopted. Ice crystals were assumed to be hexagonal columns, for IWP
$<$ 1 g/m$^2$ \citep{b20} and compact polycrystals for IWP $\geq$ 1
g/m$^2$ \citep{b7}. All particles were horizontally aligned with
random azimuthal orientation. The equal-volume-ellipsoid aspect
ratio of the particles is assumed to be 1.3.

The radiative transfer (RT) model used for this study is able to
handle also even more realistic cases, but for this first assessment
it was decided to keep the assumptions as simple as possible for
clarity.

The RT model used was the Atmospheric Radiative Transfer Simulator
(ARTS). The basics of the model are described by \citet{b3}. Here we
used version ARTS-1-1-1095, which can simulate the scattering of
radiation by cloud particles \citep{b11}. ARTS offers two different
scattering algorithms: a Monte Carlo algorithm and an iterative
discrete ordinate algorithm (DOIT). In this work we used the DOIT
algorithm, which is described in detail by \citet{b9}.

The clear-sky part of ARTS has been compared against a range of
other microwave radiative transfer models \citep{b24} and against
co-located AMSU data and radiosonde profiles \citep{b2}. The
scattering part of ARTS has been compared against several other
scattering models \citep{b8,b18}, against co-located AMSU data and
mesoscale weather prediction model fields \citep{b29}.

ARTS can handle all four Stokes components. However, the azimuthally
symmetric geometry in this case implies that only I and Q Stokes
components are non-zero. The component I represents the total
intensity (sum of horizontally and vertically polarised
intensities), the component Q represents the linear polarisation
difference.
\begin{table*}
 \vspace*{30pt}
 \centering
\caption{Simulated polarisation signal (Q) per unit IWP at three
different zenith angles. For every angle the total signal and the
polarisation induced on 2.7 K CMB are quoted.}
  \begin{tabular}{@{}ccccccc@{}}
  \hline
Frequency  &  \multicolumn{2} {c}{Zenith angle} & \multicolumn{2}{c} {Zenith angle} & \multicolumn{2} {c}{Zenith angle}\\
(GHz)  &  \multicolumn{2} {c}{5$^\circ$} & \multicolumn{2}{c}{25$^\circ$} & \multicolumn{2}{c}{45$^\circ$}\\
 \hline
     & Q$_{total}$ & Q$_{CMB}$ & Q$_{total}$ & Q$_{CMB}$ & Q$_{total}$ &  Q$_{CMB}$\\
       &  \multicolumn{2} {c}{${\mu}K/g{\cdot}m^{-2}$} & \multicolumn{2}{c}{${\mu}K/g{\cdot}m^{-2}$} & \multicolumn{2}{c}{${\mu}K/g{\cdot}m^{-2}$}\\
 \hline
 97 & 7.3 & 3.1 & 96 & 48 & 330 & 190 \\
 150 & 18 & 8.5 & 240 & 125 & 790 & 370 \\
 220 & 40 & 20 & 520 & 260 & 1700 & 850 \\
 \hline
\end{tabular}
\end{table*}

\begin{figure*}
 \vspace*{0pt}

    \includegraphics [width=18 cm]{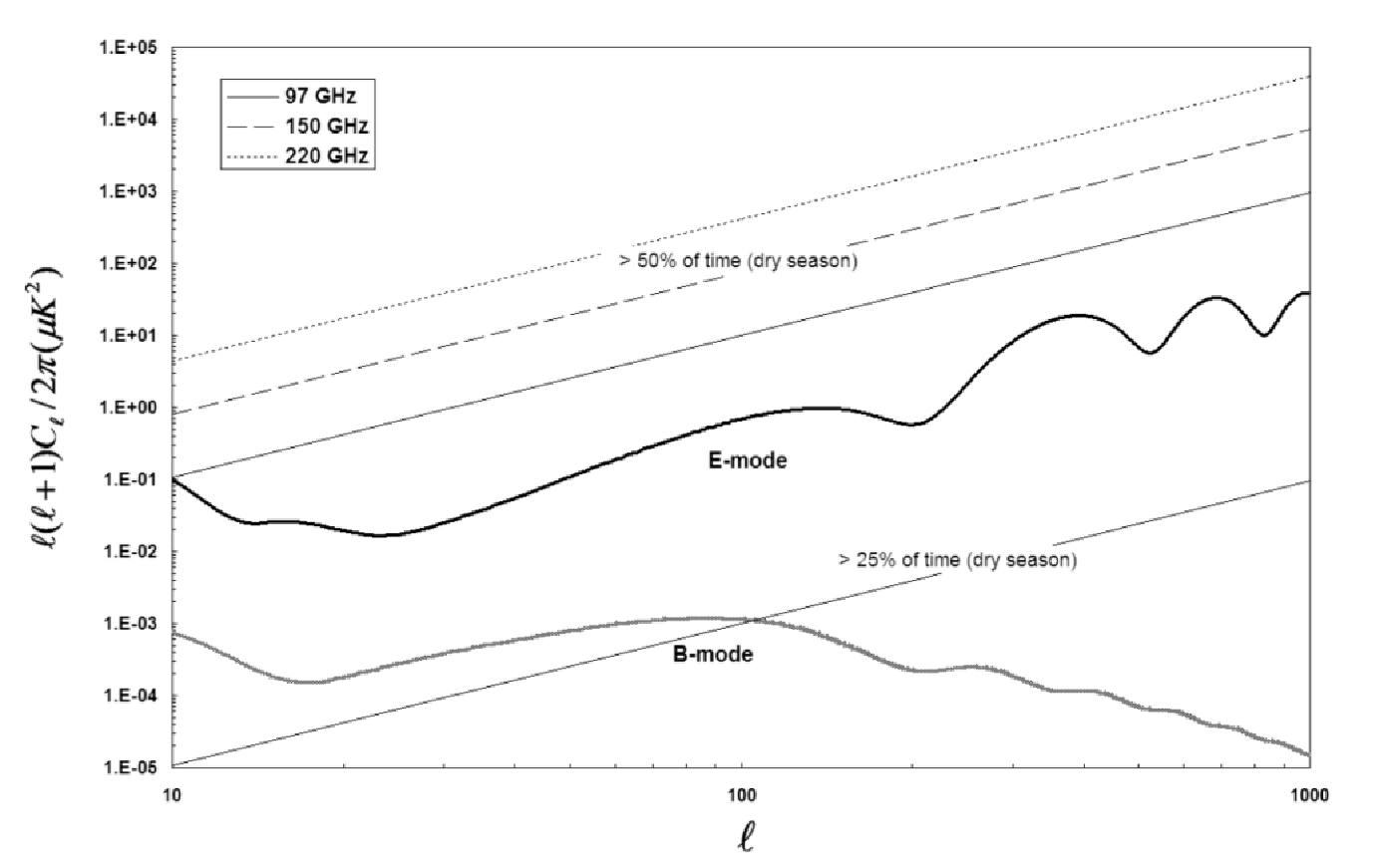}\\
    \caption{Angular power spectrum of CMB polarisation (E-mode and B-mode)
  as calculated by CMBFAST code. Straight lines represent an upper limit on the amount
  of polarization induced by ice crystal clouds on the 2.7 K CMB for more than 50\%
  observing time during dry season in Atacama (IWP = 0.01 g/m$^2$)
  at the three C$_{\ell}$OVER frequencies and an upper limit for
  25\% observing time during dry season (IWP = 0.0001 g/m$^2$) at
  97 GHz. A flat power spectrum is assumed; although it appears that the ice crystal
  signal might be dominating ground-based observations of CMB
  polarization of E and B-modes, it must be stressed that, while the CMB signal is
  fixed in the sky, the ice signal is most probably variable with time.
  Therefore it is always possible to disentangle and therefore greatly reduce
  the ice signal from the sky signal by a properly designed observing strategy}
\end{figure*}

\section{Results}

Figure 2 shows the ARTS simulation results. The sensor was assumed
to be at an altitude of 5080 m. The cloud was assumed to be located
at an altitude of 9000-11000 m. Each of the plots includes three
curves corresponding to the central frequencies of the
C$_{\ell}$OVER bands. The assumed IWP value is IWP = 1 g/m$^2$. The
different rows show the simulated radiance (top row), the radiance
difference (i.e. difference between the radiance from the cloud and
the clear sky, middle row) and the polarisation difference (Stokes
component Q, bottom row). All signals are shown as a function of the
instrument angle relative to the zenith direction. The polarisation
signal is zero at the zenith, and increases with zenith angle.

Both the radiance difference and the polarisation signal increase
with the IWP. This is shown more clearly in Figure 3, where the
simulated radiance, the radiance difference and the polarisation
difference are plotted as a function of IWP for a zenith angle of
5$^\circ$. The figure covers values of IWP ranging from 0.01 to 100
g/m$^2$. Note that the radiative transfer is in the linear regime at
these frequencies for reasonable IWP values. (Note the logarithmic
x-scale of the plots).

\section{Effect on observed CMB polarisation}

The results of the calculations using ARTS show the expected
polarisation signal induced by the ice clouds. The signal represents
the difference between a cloudy and clear sky. Cloud parameters are
derived from climatological data.

This signal is divided into two components:

\begin{enumerate}
  \item upwelling radiation, emitted by ground and lower atmosphere, that is
back-scattered and polarised by the ice crystal clouds;
  \item CMB radiation that is forward-scattered by the ice with a variation
in its polarisation (ARTS model assumes that CMB radiation is
unpolarised).
\end{enumerate}

In order to disentangle the two components, we run the ARTS model
alternately imposing T$_{ground}$ and T$_{CMB}$ equal to 0
$^\circ$K. Table 2 shows the polarisation signal per unit IWP, which
is applicable in the linear regime (at least up to 100 g/m$^2$).

In order to compare the ice crystal induced polarization with the
expected cosmological signal, we generated the angular power
spectrum of the CMB polarisation (both E-mode and B-mode) in an ACDM
(Adiabatic Cold Dark Matter) cosmology scenario with tensor to
scalar ratio r = 0.01 (using CMBFAST code, see \citet{b28}).

Figure 4 reports the expected CMB polarisation signal power spectrum
in terms of the spherical harmonic coefficient ${\ell}$. The range
is limited to spatial scales which are relevant for C$_{\ell}$OVER
experiment: between 10$^\circ$ and 10 arcmin (${\ell}$ ranging from
20 to 1000).

The polarisation signal on CMB radiation due to ice crystals is
plotted for each C$_{\ell}$OVER frequency, assuming an observing
zenith angle of 5$^\circ$ and  IWP values corresponding to an upper
limit for respectively 50\% and 25\% of observing time during dry
season.

In the figure we arbitrarily assumed a flat power spectrum for the
observed signal from ice crystal clouds; it has to be underlined
that this model is really inadequate and a more realistic
representation should take into account the cirrus cloud morphology
and spatial distribution which unfortunately are very poorly known.
Qualitatively, a decrease at the high ${\ell}$ is expected, since
the clouds are supposed to be quite homogeneous at these angular
scales, but, in any case, the signal could still be orders of
magnitude stronger that the expected B-mode component during a
significant fraction of the observing time.

\section{Conclusions}

The polarisation signal on CMB radiation due to ice crystal clouds
cannot be neglected by a ground based experiment looking for mapping
E- and B-mode patterns even for low ice water column density values
(0.001 g/m$^2$). The effect is particularly strong for the high
frequencies commonly used for CMB measurements (150 GHz and above).

Possible mitigation strategies include:

\begin{enumerate}
  \item constant elevation fast instrument scanning: at the moment C$_{\ell}$OVER
  is designed to internally modulate the CMB polarisation signal and therefore an
  absolute measurement of this parameter (i.e. for each pixel) would be possible.
  A differential approach, based on rapidly measurement of signal differences between
  contiguous pixels would cancel the constant bias due to ice crystals clouds.
  However, the measurement would still be affected by the cloud spatial distribution
  and inhomogeneities;
  \item Ice Water Path could be measured by processing Earth Observing (EO) satellite data.
  In principle geo-stationary satellites near real-time monitoring capabilities (one measurement
   every  15-30 minutes). However, as discussed above (para. 2), very low tropospheric water vapor
  content (both in Atacama and in Antarctica) seriously undermine this
  measurement. Future EO mission dedicated to ice cloud measurements
  will be helpful for solving this problem;
  \item IWP in situ measurements can be carried out in order to
  assess the quality of measured polarisation, and to characterize,
  during clean nights, a set of reference pixels to be used for ice
  detection. Among the possible options, the use of a polarised
  lidar, though difficult to deploy and operate, would provide us with a
  full characterization of ice crystals (shape, size distribution
  and orientation) allowing an accurate modelling of radiative
  effects in the microwaves.
\end{enumerate}

In this difficult observational context, a preliminary comparison
between the sites can be attempted. The presence of the sun at high
elevation angles penalizes Atacama with respect to Antarctica sites;
however, due to the high sensitivities required for CMB polarisation
measurements, it is in any case difficult to carry out measurements
with the sun above the horizon (strong signal in the instrument
sidelobes). On the other hand climatological data on ice cloud
occurrence and density suggest that Atacama observing conditions
during dry seasons are significantly better than Antarctica (with
frequent occurrences of  IWP $\leq$ 0.0001 g/m$^2$)

\section*{Acknowledgments}

Thanks to the ARTS radiative transfer community, many of whom have
indirectly contributed by implementing features to the ARTS model.

\bsp

\label{lastpage}

\end{document}